\documentclass{PoS}

\usepackage{varioref}
\usepackage[centertags]{amsmath}

\title{A Scan for Models of Neutrino Mixing from Non-Abelian Discrete Symmetries}

\ShortTitle{A Scan for Models of Neutrino Mixing}

\author{\speaker{Ak\i{}n Wingerter}\\
                        Laboratoire de Physique Subatomique et de Cosmologie, 
                        UJF Grenoble 1, CNRS/IN2P3, INPG, 53 Avenue des Martyrs, 
                        F-38026 Grenoble, France\\
        E-mail: \email{akin@lpsc.in2p3.fr}}

\author{Krishna Mohan Parattu\\
        Inter-University Centre for Astronomy and Astrophysics, Ganeshkhind, Pune 411007, India\\
        E-mail: \email{krishna@iucaa.ernet.in}}

\abstract{The structure of the neutrino mixing matrix is indicative of an underlying family symmetry that interrelates the three generations of fermions in the Standard Model. We systematically scan the parameter space of 76 discrete non-Abelian family symmetries and construct all models with the Standard Model particle content and up to three flavon fields where we include non-renormalizable interactions of mass dimension five and six. We find that of the 76 groups that we considered, 44 groups can accommodate models that are consistent with experiment at $3\sigma$, and 38 groups can have models that are tribimaximal. One immediate consequence is that $A_4$ is not ``special'', but should be considered on equal footing with other groups such as $T_7$ that is the smallest group for which we find tribimaximal mixing, and $T_{13}$ that has the largest fraction of TBM models. We present the details of a model with $\theta_{12}=33.9^\circ, \theta_{23}=49.1^\circ, \theta_{13}=5.1^\circ$ to show that a non-zero $\theta_{13}$ can easily be accommodated.}

\FullConference{The 2011 Europhysics Conference on High Energy Physics-HEP 2011,\\
		July 21-27, 2011\\
		Grenoble, Rh\^{o}ne-Alpes France}

\labelformat{section}{Section #1} 
\labelformat{subsection}{Section #1} 
\labelformat{equation}{Eq.~(#1)} 
\labelformat{figure}{Fig.~#1} 
\labelformat{subfigure}{Fig.~\thefigure#1} 

\newcommand{\SU}[1]{\ensuremath{\mathrm{SU}(#1)}}

\newcommand{\code}[1]{{\small{\texttt{#1}}}}
\newcommand{\gapid}{\code{GAPID}}
\newcommand{\gap}{\code{GAP}}

\newcommand{\UPMNS}{\ensuremath{U_\mathrm{PMNS}}}

\newcommand{\vevs}{{\small{\textit{vevs}}}}
\newcommand{\gid}[2]{\ensuremath{\mathfrak{G}({#1},{#2})}}
\newcommand{\bs}[1]{\ensuremath{\boldsymbol{#1}}}
\newcommand{\bsn}[2]{\ensuremath{\boldsymbol{#1^{(#2)}}}}

\begin{document}

\section{Introduction}
\label{sec:introduction}

The discovery of neutrino masses is one of the most exciting developments in physics during the last decade. Unlike the CKM matrix in the quark sector, the mixing matrix \UPMNS{} contains large angles, and until very recently its form was well-approximated by a pattern that was subsequently dubbed tribimaximal \cite{Harrison:2002er}. The results from the T2K \cite{Abe:2011sj} and MINOS \cite{Adamson:2011qu} collaborations have called this paradigm into question, and global fits to neutrino oscillation data now indicate that $\theta_{13}\gneqq0$ with about $3\sigma$ significance \cite{Schwetz:2011zk}. 

The (near) tribimaximal mixing pattern has given rise to speculation that there might exist a symmetry that interrelates the three generations of leptons in the Standard Model. The transformation properties of the leptons under this so-called family symmetry then determine the Lagrangian and ultimately the mixing pattern. For a recent review on finite symmetries and their application to neutrino physics, see Ref.~\cite{Grimus:2011fk}. In the past, models based on $S_3$, $D_4$, $A_4$, $S_4$, $A_5$, $T'$, $\Delta(27)$, to name a few, have been particularly popular, and it has been argued that $A_4$ is inherently connected to tribimaximal mixing (TBM). Here we describe a systematic approach to the problem of finding the family symmetry that may best describe the observed pattern of mixing angles \cite{Parattu:2010cy}. To that end, we use \gap{} \cite{GAP4} for the representation theory of discrete groups and an algorithm due to van den Broek and J.F. Cornwell \cite{vandenbroek:1978aa} to calculate the Clebsch-Gordan coefficients for any finite group. The availability of such powerful algebraic tools and the fact that the Lagrangian is fully determined by the underlying symmetries of the theory render the calculation of neutrino mixing angles an ideal problem for a computational approach.

\section{Algorithm for the Scan}
\label{sec:scan}

There are 1,048 discrete groups with less than or equal to 100 elements. Since we want to interrelate the three generations of leptons, we will put the lepton doublets $L_{e}$, $L_{\mu}$, $L_{\tau}$ into one and the same irreducible triplet representation of the family symmetry. This leaves us with 90 groups that have three-dimensional irreducible representations. It turns out that for 14 groups a comprehensive scan would take too long, so in the end we are left with 76 groups \cite{Parattu:2010cy}. For the right-handed leptons $e$, $\mu$, $\tau$ and the two Higgs doublets $h_u$ and $h_d$, we assume that they transform in one-dimensional (not necessarily singlet) representations. The choice for the Higgs fields is motivated by supersymmetry, whereas the assumptions for the right-handed leptons should eventually be generalized. We spontaneously break the family symmetry by giving vacuum expectation values (\vevs{}) to flavon fields $\varphi_T$, $\varphi_S$, $\xi$ that are gauge singlets, but may transform in any representation of the family symmetry. We include in the Lagrangian all terms of mass dimension less than or equal to 6. In scanning over the \vevs{} for the flavon fields we have not been fully general, but restricted ourselves to those with entries 0 or 1 (times a constant). We agree that we are sampling only a discrete part of the parameter space, but choosing continuous values is computationally not viable. Moreover, that the \vevs{} are all of the same order of magnitude is not an unreasonable assumption.

\section{Discussion of Results}
\label{sec:results}

\begin{figure}[h!]
\addtolength{\abovecaptionskip}{-1mm}
\includegraphics[width=1.0\textwidth]{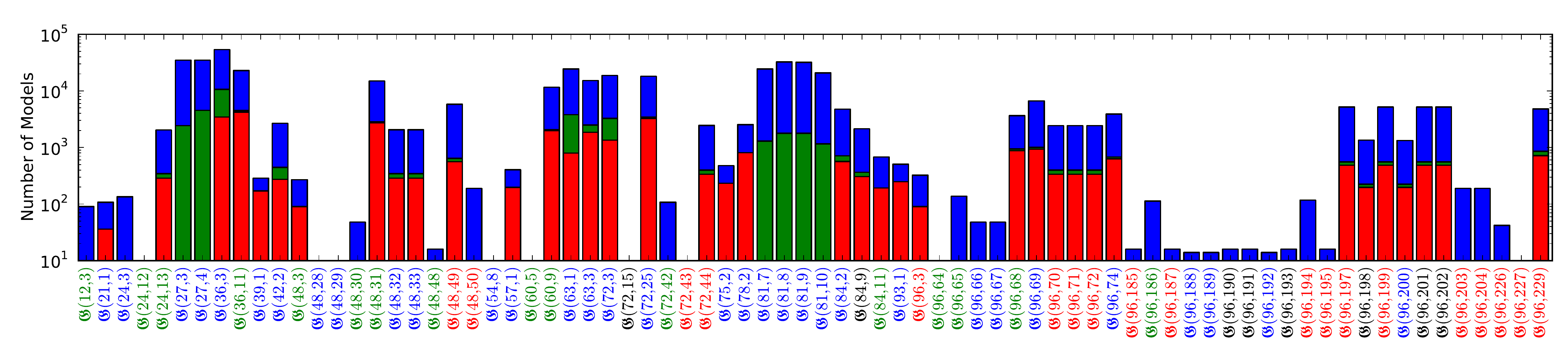}
\caption{The number of models per symmetry group. The green and the red bars indicate the number of models that for given \vevs{} lie within the $3\sigma$ interval or are tribimaximal mixing, respectively.}
\label{fig:numlagtbm}
\end{figure}

In \ref{fig:numlagtbm}, we summarize the results of our scan. Not all of the groups that we have considered have names in the conventional sense, so in the following we will denote them by their so-called \gapid{}s \cite{Parattu:2010cy}. E.g.~\gid{12}{3} is the alternating group $A_4$, i.e.~it is the third in the list of all groups of 12 elements. Out of the 76 groups that we have scanned, 9 (12$\%$) have only singular mass matrices. 44 groups (58$\%$) lie in the 3$\sigma$ interval\footnote{Note that in this section we refer to the experimental limits that we used in our original work \cite{Parattu:2010cy}.}, and 38 (50$\%$) are even tribimaximal (for at least one vacuum configuration, respectively). We find that the smallest group that can realize TBM is  $T_7$, and the one with the largest fraction of TBM models is $T_{13}$. Regarding whether there is a special connection between $A_4$ and TBM, we find that taking the fraction of TBM models as a criterion, $T_7$ and $T_{13}$ are as promising, if not more, than $A_4$. Furthermore, only 16 of the 35 groups containing $A_4$ as a subgroup (red and green labels on the $x$-axis) realize TBM, while there are 22 groups which do not have $A_4$ as a subgroup but do realize TBM.

\begin{figure}[h!]
\addtolength{\abovecaptionskip}{-9mm}
\centering
\includegraphics[width=1.0\textwidth]{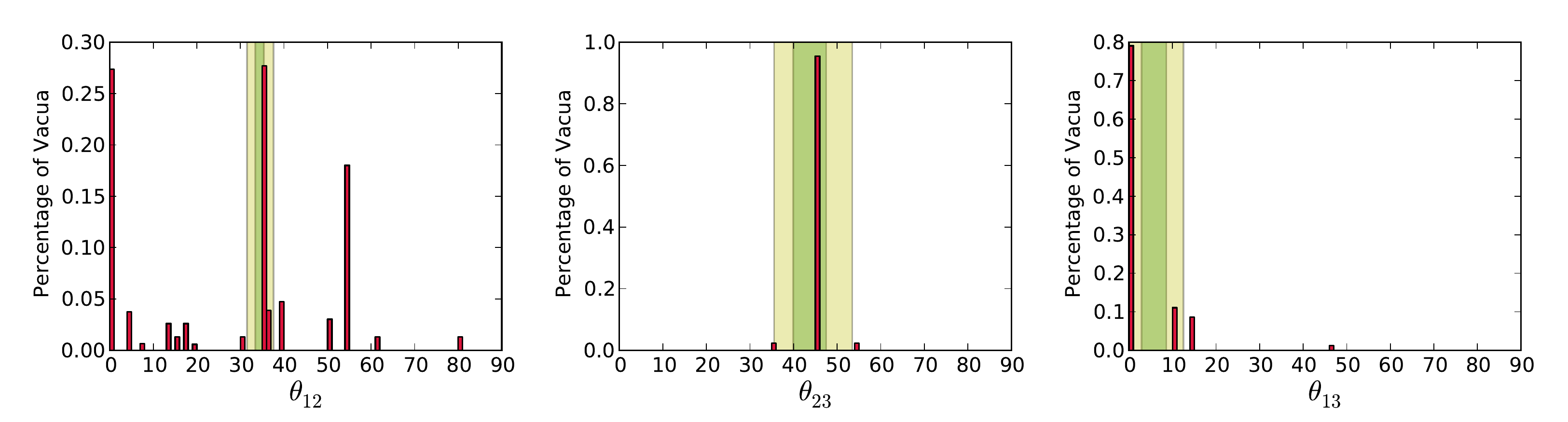}
\label{fig:g39-1-angles-histo-1d-b}
\caption{Number of vacua with family symmetry $\gid{39}{1}=T_{13}$ that give the mixing angles on the $x$-axis.}
\label{fig:g39-1-angles-histo-1d}
\end{figure}

In \ref{fig:g39-1-angles-histo-1d}, we show the distribution of the mixing angles for one particular group, namely $T_{13}$. In the first histogram we count the number of vacua that realize $\theta_{12}$ as given on the $x$-axis, where $\theta_{23}$ and $\theta_{13}$ are restricted to their $3\sigma$ intervals. Analogous statements hold for the two other histograms. The maxima in all three histograms correspond to tribimaximal mixing, and we obtain the almost unique prediction $\theta_{23}=45^\circ$. The most preferred value for $\theta_{13}$ is $0^\circ$, but non-zero values are also allowed in the $3\sigma$ band.

\section{A Model with $\bs{\theta_{13} \neq 0}$}
\label{sec:model}


We present a specific model \cite{Parattu:2010cy} where $\theta_{13} \neq 0$ and all mixing angles are compatible with experiment \cite{Schwetz:2011zk} at $2\sigma$.  For the family symmetry, we take $\gid{36}{11}=A_4\times \mathbb{Z}_3$ and assign the following family charges to the Standard Model particles and flavon fields:
\begin{equation}
(L,e,\mu,\tau,h_u,h_d,\varphi_T,\varphi_S,\xi) \sim (\bs{3},\bs{1},\bs{1'''},\bsn{1}{4},\bs{1'},\bsn{1}{7},\bsn{1}{5},\bs{3''},\bs{3''}).
\end{equation}
The Lagrangian is given by all the terms that are invariant under both the gauge and the family symmetries and are at most of mass dimension 6:
\begin{align*}
c_T\, L L h_u h_u \varphi_T + c_S\, L L h_u h_u \varphi_S + &c_\xi\, L L h_u h_u \xi + d_{S}^{(e)}\, L e h_d \varphi_S + d_{\xi}^{(e)}\, L e h_d \xi \\
&+ d_{S}^{(\mu)}\, L \mu h_d \varphi_S + d_{\xi}^{(\mu)}\, L \mu h_d \xi + d_{S}^{(\tau)}\, L \tau h_d \varphi_S + d_{\xi}^{(\tau)}\, L \tau h_d \xi
\end{align*}
Here, we have only listed those terms that will contribute to the mass matrix. The $c$'s and $d$'s are some coefficients that can be related to the Yukawa couplings in a basis where the charged lepton mass matrix is diagonal. After contracting the family and \SU{2} gauge indices and giving \vevs{} to the flavon fields along the directions $\langle\varphi_{T}\rangle = v_T$, $\langle\varphi_{S}\rangle = (v_S,0,v_S)$, $\langle\xi\rangle = (v_\xi, v_\xi, v_\xi)$, we obtain the mixing angles $\theta_{12}=33.9^\circ$, $\theta_{23}=49.1^\circ$ and $\theta_{13}=5.1^\circ$ that all lie in the $2\sigma$ interval of the experimentally determined values. We have assumed that all \vevs{} are of the same order of magnitude (the absolute value of this scale does not enter the calculation of the mixing angles).

\section{Conclusions \& Outlook}

In this work, we pursued a systematic search for models of neutrino flavor that reproduce the observed mixing pattern. We find that half of the 76 flavor groups that we considered can give rise to tribimaximal mixing. As a consequence, we cannot affirm that $A_4$ is ``special'', in particular as compared to $T_7$ or $T_{13}$. Following the recent hints of a non-vanishing $\theta_{13}$ we presented a model where all mixing angles are compatible with experiment at $2\sigma$. In a next step, we will extend our analysis to include neutrino masses and the stabilization of the flavon vacuum expectation values.


\bibliography{mybibliography}

\bibliographystyle{./JHEP}

\end{document}